\input harvmac
\tolerance=10000



%
\def\eql{~=~}

\def\al{\alpha}

\def\coeff#1#2{\relax{\textstyle {#1 \over #2}}\displaystyle}

\def\gop{\mathop{g}^{\!\circ}{}}

\def\tobe{IIB}
\def\none{\cN\!=\!1}
\def\ntwo{\cN\!=\!2}
\def\nfour{\cN\!=\!4}
\def\neigh{\cN\!=\!8}
\def\cA{{\cal A}} \def\cB{{\cal B}}

\def\cJ{{\cal J}} 
 
\def\cN{{\cal N}} \def\cO{{\cal O}}
\def\cP{{\cal P}} 
\def\cR{{\cal R}}

\def\bfone{\relax{\rm 1\kern-.35em 1}}
\def\inbar{\vrule height1.55ex width.4pt depth0pt}
\def\IC{\relax\,\hbox{$\inbar\kern-.3em{\rm C}$}}
\def\ID{\relax{\rm I\kern-.18em D}}
\def\IF{\relax{\rm I\kern-.18em F}}
\def\IH{\relax{\rm I\kern-.18em H}}
\def\II{\relax{\rm I\kern-.17em I}}
\def\IN{\relax{\rm I\kern-.18em N}}
\def\IP{\relax{\rm I\kern-.18em P}}
\def\IQ{\relax\,\hbox{$\inbar\kern-.3em{\rm Q}$}}
\def\IR{\relax{\rm I\kern-.18em R}}
\def\us{\bf}
\font\cmss=cmss10 \font\cmsss=cmss10 at 7pt
\def\ZZ{\relax\ifmmode\mathchoice
{\hbox{\cmss Z\kern-.4em Z}} {\hbox{\cmss Z\kern-.4em Z}}
{\lower.9pt\hbox{\cmsss Z\kern-.4em Z}}
{\lower1.2pt\hbox{\cmsss Z\kern-.4em Z}}\else{\cmss Z\kern-.4em Z}\fi}

\def\gop{\mathop{g}^{\!\circ}{}}
\def\sech{{\mathop{\rm sech}}}
\def\nihil#1{{\it #1}}
\def\eprt#1{{\tt #1}}
\def\nup#1({Nucl.\ Phys.\ $\us {B#1}$\ (}
\def\plt#1({Phys.\ Lett.\ $\us  {#1B}$\ (}
\def\cmp#1({Comm.\ Math.\ Phys.\ $\us  {#1}$\ (}
\def\prp#1({Phys.\ Rep.\ $\us  {#1}$\ (}
\def\prl#1({Phys.\ Rev.\ Lett.\ $\us  {#1}$\ (}
\def\prv#1({Phys.\ Rev.\ $\us  {#1}$\ (}
\def\mpl#1({Mod.\ Phys.\ Let.\ $\us  {A#1}$\ (}
\def\ijmp#1({Int.\ J.\ Mod.\ Phys.\ $\us{A#1}$\ (}
\def\atmp#1({Adv.\ Theor.\ Math.\ Phys.\ $\bf {#1}$\ (}
\def\cqg#1({Class.\ Quant.\ Grav.\ $\bf {#1}$\ (}
\def\jag#1({Jour.\ Alg.\ Geom.\ $\us {#1}$\ (}
\def\jhep#1({JHEP $\bf {#1}$\ (}

%


%
\lref\JMalda{J.~Maldacena, \nihil{The Large $N$ Limit of Superconformal
Field Theories and Supergravity,}, Adv.~Theor. Math. Phys.~{\bf 2}
(1998) 231 \eprt{hep-th/9711200}.}
%
\lref\NSsixteen{
N.~Seiberg, \nihil{Notes on theories with 16 supercharges,}
Nucl.\ Phys.\ Proc.\ Suppl.\  {\bf 67} (1998) 158;
\eprt{hep-th/9705117}.}
%
\lref\NDorey{
N.~Dorey, \nihil{An elliptic superpotential for softly broken N = 4
supersymmetric   Yang-Mills theory,}  \jhep{9907}  (1999) 021;
\eprt{hep-th/9906011}.
}
%
\lref\OANDSPK{
N.~Dorey and S.~P.~Kumar, \nihil{Softly-broken N = 4 supersymmetry
in the large-N limit,}
\jhep{0002} (2000) 006; \eprt{hep-th/0001103}.\hfill\break
O.~Aharony, N.~Dorey and S.~P.~Kumar,
\nihil{New modular invariance in the N = 1* theory, operator mixings and
supergravity singularities,} \jhep{0006} (2000) 026;
\eprt{hep-th/0006008}.}
%
\lref\DonWit{R.~Donagi and E.~Witten,
\nihil{Supersymmetric Yang-Mills Theory And Integrable Systems,}
\nup{460} (1996) 299;  \eprt{hep-th/9510101}.}
%
\lref\PolStr{J.~Polchinski and M.~J.~Strassler,
{\it The String Dual of a Confining Four-Dimensional Gauge Theory,}
\eprt{hep-th/0003136}.}
%
\lref\JSPow{J.~H.~Schwarz, \nihil{The Power of M theory,} \plt{367}
(1996)  97;  \eprt{hep-th/9510086}.}
%
\lref\PAdual{ P.~S.~Aspinwall, \nihil{Some relationships between
dualities in string theory,} Nucl.\ Phys.\ Proc.\ Suppl.\
{\bf 46}  (1996) 30;  \eprt{hep-th/9508154}.}
%
\lref\KPW{A.\ Khavaev, K.\ Pilch and N.P.\ Warner, \nihil{New
Vacua of Gauged  ${\cal N}=8$ Supergravity in Five Dimensions},
\plt{487}  (2000) 14;  \eprt{hep-th/9812035}.}
%
\lref\FGPWa{D.~Z. Freedman, S.~S. Gubser, K.~Pilch, and N.~P. Warner,
\nihil{Renormalization Group Flows from Holography---Supersymmetry
and a c-Theorem,}   \atmp{3} (1999) 363; \eprt{hep-th/9904017} }
%
\lref\KLM{A.~Karch, D.~Lust and A.~Miemiec, \nihil{New N = 1 superconformal
field  theories and their supergravity  description,} \plt{454} (1999) 265;
\eprt{hep-th/9901041}.}
%
\lref\RGLMJS{ R.~G.~Leigh and M.~J.~Strassler, \nihil{Exactly marginal
operators and duality in four-dimensional N=1 supersymmetric gauge theory,}
\nup{447} (1995)  95; \eprt{hep-th/9503121}.}
%
\lref\Ahnetal{C.~Ahn and J.~Paeng,
\nihil{Three-dimensional SCFTs, supersymmetric domain wall and
renormalization  group flow,}
\nup{595} (2001) 119;
\eprt{hep-th/0008065}.}
\lref\AhnWoo{C.~Ahn and K.~Woo, \nihil{Supersymmetric domain wall and RG
flow from 4-dimensional gauged N = 8  supergravity,}
\nup{599} (2001)  83;  \eprt{hep-th/0011121}.}
%
\lref\NPWpot{ N.~P.~Warner,
\nihil{Some Properties of The Scalar Potential in 
Gauged Supergravity Theories,}
\nup{231}  (1984) 250.}
%
\lref\NPWext{ N.~P.~Warner, \nihil{Some New Extrema of the Scalar
Potential of Gauged N=8 Supergravity,} \plt{128}  (1983) 169 .}
%
\lref\dWNic{ B.~de Wit and H.~Nicolai,
\nihil{N=8 Supergravity with Local SO(8) X SU(8) Invariance,}
\plt{108} (1982)  285;  \nihil{N=8 Supergravity,}
Nucl.\ Phys.\  {\bf B  208}, 323 (1982). }
%
\lref\BdWHNNW{ B.~de Wit, H.~Nicolai and N.~P.~Warner,
\nihil{The Embedding of Gauged N=8 Supergravity into D = 11 Supergravity,}
\nup{255} (1985)  29.}
%
\lref\ConTrunc{ B.~de Wit and H.~Nicolai, \nihil{The Consistency of the $S^7$
Truncation in D = 11 Supergravity,}
\nup{281}  (1987) 211.}
%
\lref\HNNW{ H.~Nicolai and N.~P.~Warner, \nihil{The SU(3) X U(1) Invariant
Breaking of Gauged N=8 Supergravity,} \nup{259} (1985) 412.}
%
\lref\KPNWa{K.\ Pilch and N.P.\ Warner, \nihil{A New Supersymmetric
Compactification of Chiral IIB Supergravity,} \plt{487} (2000) 22;
\eprt{hep-th/0002192}.}
%
\lref\KPNWb{K.~Pilch and N.P.~Warner, \nihil{$\cN=2$ Supersymmetric
RG Flows and the IIB  Dilaton,} \nup{594} (2001) 209;
\eprt{hep-th/0004063}.}
%
\lref\KPNWc{
K.~Pilch and N.~P.~Warner,
\nihil{N=1 Supersymmetric Renormalization Group Flows from IIB Supergravity,}
\atmp{4} (2000) 627-677, \eprt{hep-th/0006066}.}
%
\lref\KPNWntwo{K.~Pilch and N.P.~Warner, \nihil{$\ntwo$ Supersymmetric
RG Flows and the IIB  Dilaton,} \nup{594} (2001) 209;
\eprt{hep-th/0004063}.}
%
\lref\CNPNPW{C.~N.~Pope and N.~P.~Warner,
\nihil{An SU(4) Invariant Compactification of D = 11 Supergravity
on a Stretched Seven Sphere,}  \plt{150}  (1985) 352;
\nihil{Two New Classes Of Compactifications of D = 11 Supergravity,}
Class.\ Quant.\ Grav.\  \cqg{\bf 2}, L1 (1985).}
%
\lref\dWNWnew{B.~de Wit, H.~Nicolai and N.~P.~Warner,
\nihil{work in progress}.}
\lref\LJR{ L.~J.~Romans, \nihil{New Compactifications of Chiral N=2,
 D = 10 Supergravity,} \plt{153} (1985)   392. }
%
\lref\CVJprim{C.~V.~Johnson, \nihil{D-brane Primer,}
hep-th/0007170. }
%
\lref\Buscher{T.~H.~Buscher,
\nihil{A Symmetry of the String Background Field Equations,}
\plt{194}  (1987)  59;  \nihil{Path Integral Derivation of Quantum Duality
in Nonlinear Sigma Models,}
\plt{201}  (1988) 466; \nihil{Studies of the Two-Dimensional Nonlinear
Sigma-Model,}  PhD Thesis,  UMI 89-00205.}
%
\lref\AKNW{A.~Khavaev and N.~P.~Warner,
\nihil{An N=1 Supersymmetric Coulomb Flow,} CITUSC/01-021; USC-01/03;
\eprt{hep-th/0106032}.}
%
\lref\Kehagias{A.~Kehagias, \nihil{New Type IIB Vacua and their
F-theory Interpretation,} \plt{435} (1998)  337; \eprt{hep-th/9805131}.}
%
\lref\AFHS{B.~S.~Acharya, J.~M.~Figueroa-O'Farrill, C.~M.~Hull and
B.~Spence, \nihil{Branes at Conical Singularities and Holography,}
\atmp{2} (1999)  1249; \eprt{hep-th/9808014}.}
%
\lref\DPEins{C.P.~Boyer, K.~Galicki and  M.~Nakamaye,
\nihil{On the Geometry of Sasakian-Einstein 5-Manifolds,}
\eprt{math.DG/0012047}.}
\lref\ECBJ{ E.~Cremmer and B.~Julia, \nihil{The SO(8) Supergravity,}
\nup{159} (1979) 141.}
%
\lref\IKEW{I.~R.~Klebanov and E.~Witten, \nihil{Superconformal Field Theory
on Threebranes at a Calabi-Yau  Singularity,} \nup{536} (1998)  199;
\eprt{hep-th/9807080}.}
%
\lref\SGNNSS{S.~Gubser, N.~Nekrasov and S.~Shatashvili, \nihil{Generalized
Conifolds and Four Dimensional N = 1 Superconformal  Theories,}
\jhep{9905}   (1999) 003;  \eprt{hep-th/9811230}.}
%
\lref\CVJprobe{ C.~V.~Johnson, K.~J.~Lovis and D.~C.~Page,
\nihil{Probing some N = 1 AdS/CFT RG flows,}
\jhep{0105} (2001) 036;  \eprt{hep-th/0011166}.}
%
\lref\CveticTB{M.~Cvetic, H.~Lu and C.~N.~Pope,
\nihil{Geometry of the embedding of supergravity scalar manifolds 
in D = 11  and D = 10,} \nup{584} (2000) 149;
\eprt{hep-th/0002099}.}
%
\lref\CVJnew{C.V.~Johnson, K.~Lovis and D.~Page, 
\nihil{The K\"ahler Structure of Supersymmetric Holographic RG Flows,}
{to appear.} }
\lref\IKMS{
I.~R.~Klebanov and M.~J.~Strassler,
\nihil{Supergravity and a confining gauge theory: Duality cascades and  
chiSB-resolution of naked singularities,} JHEP {\bf 0008} (2000)  052; 
\eprt{hep-th/0007191}.}
%
\lref\SSGtop{S.~S.~Gubser, \nihil{Supersymmetry and F-theory realization 
of the deformed conifold with  three-form flux,}
\eprt{hep-th/0010010}.}
%
\lref\BPP{ A.~Buchel, A.~W.~Peet and J.~Polchinski,
\nihil{Gauge dual and noncommutative extension of an N = 2 supergravity 
solution,} \prv{D63} (2001)  044009; \eprt{hep-th/0008076}.}
%
\lref\EJP{
N.~Evans, C.~V.~Johnson and M.~Petrini,
\nihil{The enhancon and N = 2 gauge theory/gravity RG flows,}
JHEP {\bf 0010}  (2000) 022;  \eprt{hep-th/0008081}.}
%
\lref\GPPZ{
L.~Girardello, M.~Petrini, M.~Porrati and A.~Zaffaroni,
Nucl.\ Phys.\ B {\bf 569}  (2000) 451,
\eprt{hep-th/9909047}.}
\lref\FGPWb{
D.~Z.~Freedman, S.~S.~Gubser, K.~Pilch and N.~P.~Warner,
JHEP {\bf 0007} (2000) 038, \eprt{hep-th/9906194}.}

\Title{ \vbox{ \hbox{CITUSC/01-023} \hbox{USC-01/04} \hbox{\tt
hep-th/0107220} }} {\vbox{\vskip -1.0cm
\centerline{\hbox
{An ${\cal N} \!=\! 2$ Supersymmetric Membrane Flow}}
\vskip 8 pt
\centerline{
\hbox{}}}}
\vskip -.3cm
\centerline{Richard Corrado, Krzysztof Pilch and Nicholas P.\ Warner }
\medskip
\centerline{{\it Department of Physics and Astronomy}}
\centerline{{\it and}}
\centerline{{\it CIT-USC Center for
Theoretical Physics}}
\centerline{{\it University of Southern California}}
\centerline{{\it Los Angeles, CA 90089-0484, USA}}

\bigskip
\bigskip
We find $M$-theory solutions  that are holographic duals
of flows of the maximally supersymmetric ($\cN=8$) scalar-fermion
theory in $(2+1)$ dimensions.  In particular, we construct the
$M$-theory solution dual to a flow in which a single chiral multiplet 
is given a mass,  and the  theory goes to a new infra-red fixed point.  
We also examine this new solution using $M2$-brane probes.
The $(2+1)$-dimensional field theory fixed-point is closely related to that 
of Leigh and Strassler, while the $M$-theory solution is closely
related to the corresponding \tobe\ flow solution.  We recast the
\tobe\ flow solution in a more geometric manner and use this
to obtain an Ansatz for the $M$-theory flow.  We are able to
generalize our solution further to obtain flows with del Pezzo
sub-manifolds, and we give an explicit solution with a 
conifold singularity.

\vskip .3in
\Date{\sl {July, 2001}}

\parskip=4pt plus 15pt minus 1pt
\baselineskip=15pt plus 2pt minus 1pt

\newsec{Introduction}

Holographic RG flows have been fairly widely studied in using $D3$
branes in \tobe\ supergravity, but considerably less has been done for
large $N$ theories on branes of other dimensions.  There are several
fairly obvious reasons for this, but probably the primary reason is
that \tobe\ supergravity is dual, on the $D3$-brane, to a very
interesting theory: $\nfour$ supersymmetric Yang-Mills. Moreover, one
can study flows of this theory in which some, or all of the
supersymmetry is broken, and the resulting field theory on the brane
exhibits some very non-trivial quantum behavior.  The other maximally
supersymmetric holographic field theories proposed in \JMalda\ arise
in M-theory, and correspond to superconformal theories on either a
stack of $M5$-branes or on a stack of $M2$-branes.  Here we focus on
the latter, and this is primarily because the theory on the brane is a
renormalizable, $(2+1)$-dimensional field theory that is closely
related to $\nfour$ Yang-Mills theory in $(3+1)$ dimensions.

The field theory on the worldvolume of a single $M2$-brane is a
conformally invariant $\neigh$ supersymmetric theory (16
supersymmeties) with eight scalar fields and eight fermions. As
required by superconformal invariance in three-dimensions, there is an
$SO(8)$ $\cR$-symmetry under which the scalars, fermions and
supersymmetries transform as the ${\bf 8}_v$, ${\bf 8}_c$ and ${\bf
8}_s$, respectively.  On a collection of $(N+1)$ $M2$-branes, there is
an $SO(8)$-invariant theory with $8(N+1)$ scalar degrees of freedom,
corresponding to the transverse positions of $(N+1)$ $M2$-branes, as
well as their superpartners. One of these $\neigh$ multiplets
corresponds to the free theory describing the center-of-mass motion
of the system and is decoupled. The remaining degrees of freedom
parameterize a moduli space $(\IR^8)^N/S_{N+1}$. At the fixed points
in the moduli space, the theory is an interacting superconformal field
theory~\NSsixteen.  This field theory also arises as a UV limit of the
Kaluza-Klein reduction of $\nfour$ supersymmetric Yang-Mills theory on
a circle. The extra scalars in three-dimensions come from the
components of the gauge fields along the circle and a Wilson line
parameter around the circle.

In supergravity, or M-theory, the maximally supersymmetric solution
dual to the large $N$, brane vacuum configuration is the
compactification of M-theory on $AdS_4 \times S^7$.  There is a
consistent truncation \refs{\ConTrunc,\BdWHNNW} of this supergravity
theory to the massless sector, that is, to gauged $\neigh$
supergravity in four dimensions \dWNic.  This gauged supergravity theory in
four dimensions contains 70 scalar fields, and these are
holographically dual to the (traceless) bilinears in the scalars and
fermions:
\eqn\bilinears{\eqalign{\cO^{IJ} ~=~ & \Tr~\big(X^I \, X^J) ~-~ \coeff{1}{8}\,
\delta^{IJ}\, \Tr~\big(X^K \, X^K\big) \,, \quad I,J, \dots =1,\dots,8\cr
\cP^{AB} ~=~ & \Tr~\big(\lambda^A \, \lambda^B\big) ~-~ \coeff{1}{8}\,
\delta^{AB}\, \Tr~\big(\lambda^C \, \lambda^C \big) \,, \quad A,B,
\dots =1,\dots,8\,,}}
where $\cO^{IJ}$ transforms in the ${\bf 35}_v$  of $SO(8)$, and $\cP^{AB}$ 
transforms in the ${\bf 35}_c$.
Thus,  gauged $\neigh$ supergravity in four dimensions can be used to study
mass perturbations, and a uniform subsector of the  Coulomb branch of
the $\neigh$ field theory on the $M2$-brane.  The gauged
$\neigh$ supergravity in four dimensions thus plays a very analogous
role to the  gauged $\neigh$ supergravity in five dimensions.
There is, however, a significant difference:  the Yang-Mills theory
has a freely choosable (dimensionless) coupling constant and
$\theta$-angle, and these are dual to a pair of scalars
in the five-dimensional gauged supergravity theory.  The scalar-fermion
theory on the $M2$ branes has no free coupling: In particular,
three-dimensional  super Yang-Mills theory flows to the UV interacting 
superconformal fixed point, the gauge coupling is driven to 
infinity~\NSsixteen. There are thus no supergravity fields dual to a 
coupling: there are only masses and vevs in the dual of the four-dimensional 
gauged supergravity.

The fact that the conformal scalar-fermion theory is necessarily strongly
coupled makes it, {\it a priori}, hard to analyze.   However, we understand
its holographic dual as well as we understand that of $\nfour$ Yang-Mills
theory.   More to the point, the general (non-conformal)
$\neigh$ scalar-fermion theory has played
a very interesting role in helping us understand softly broken
$\nfour$ Yang-Mills, that is the so-called $\none^*$ theories.
In a beautiful paper \NDorey\ Dorey argued that one could compute
a quantum exact superpotential for the softly broken scalar-fermion theory
considered as a reduction of softly broken $\nfour$ Yang-Mills on
a circle.  Even more remarkably, it was argued that this superpotential
was independent of the radius of the circle, and thus gave quantum exact
information about ground states and domain walls of the original
$\nfour$ Yang-Mills theory.   This contention was strongly supported by
the fact that the superpotential exactly reproduced the known
quantum ground states structure of the $\none^*$ theories \DonWit.
The role of the modular $S$-duality group was manifest in the
results of \NDorey, and was subsequently used to great effect in
\OANDSPK\ to obtain exact modular expressions for various ground-state
vevs.  This also proved to be a powerful tool in probing the
brane description of the $\none^*$ flows of Yang-Mills theory
\refs{\GPPZ,\PolStr,\KPNWc,\OANDSPK}.

In terms of branes, the link between the scalar-fermion theory
and Yang-Mills theory can be implemented rather directly: One simply
compactifies $D3$-branes on a circle and T-dualizes.  The
result is a uniform distribution of $D2$-branes on the
T-dual circle in IIA supergravity.  This can then be lifted to a
distribution of $M2$ branes on a torus in M-theory.  A distribution
of such branes can then be analyzed in terms of a Coulomb branch
deformation of a set of localized $M2$-branes.
If one considers  a set of $D3$-branes in the large $N$ limit
then, because of the cosmological factor, the radius of the
compactifying circle  becomes very large in the UV,
and very small in the IR.  The fact that the radius of the circle
vanishes in the IR means that the one {\it must work}
with the T-dual description to understand properly the infra-red
behavior.  In this paper we will examine the link between the
large $N$ Yang-Mills and the large $N$ scalar-fermion theory
by looking more closely at links between the
holographically dual theories.  We will do this in a context
that is not part of the $\none^*$ flows of 
\refs{\GPPZ,\PolStr,\NDorey,\OANDSPK},
but instead will look at the M-theory analog of the Leigh-Strassler
fixed point \RGLMJS.

The dual supergravity theories are the gauged $\neigh$ supergravity
theories in five (for $D3$ branes) and four (for $M2$ branes)
dimensions.  The scalar fields of these theories live on
an $E_{6(6)}/USp(8)$ and an $E_{7(7)}/SU(8)$ coset respectively.
These groups will play a significant role in this paper, and it is
important to understand their interrelationship.  First,
$E_{6(6)}$ commutes with an $SO(1,1)$ in $E_{7(7)}$.  This extra
non-compact scalar may be  identified with the relative scale of
of the compactifying $S^5$ and the circle upon which
the $D3$ branes are wrapped.  The group, $E_{6(6)}$,
contains a particularly important subgroup, $SL(6,\IR) \times
SL(2,\IR)$.    The non-compact scalars in the $SL(6,\IR)$ lie in
the ${\bf 20'}$ of $SO(6)$ and are holographically dual to scalar bilinears
in the $\nfour$ Yang-Mills theory.  If one turns on scalars in the
${\bf 20'}$ only, the round $S^5$ upon which the \tobe\  theory is
compactified is deformed ellipsoidally to the surface:
\eqn\squish{\big\{x \in \IR^P : \ x^T\, S^T S \, x ~=~1 \big\} \ \,,}
where $P=6$ and  $S \in SL(6,\IR)$.  The non-compact scalars in
$SL(2,\IR)$
correspond to the gauge coupling and $\theta$-angle, and are thus
moduli of the supergravity theory as well.

The situation is similar, but different for $M2$ branes.  The
group $E_{7(7)}$ has a maximal non-compact subgroup $SL(8,\IR)$.
This group contains the abovementioned $SL(6,\IR) \times  SL(2,\IR)$
in the obvious manner.  Indeed the latter commutes with an $SO(1,1)$
generator defined in $SL(8,\IR)$ by:
\eqn\SOoneone{h ~\equiv~ {\rm diag}(1,1,1,1,1,1,-3,-3)  \,.}
Moreover, the subgroup of $E_{7(7)}$ that commutes with $h$
is precisely $E_{6(6)}$.  While the embedding of the groups is
extremely straightforward, the gauged supergravity theories
do not embed directly into one another.  This is because the
minimal couplings are rather different ($SO(6)$ {\it vs.} $SO(8)$),
and as a result the supersymmetrization proceeds differently, and in
particular the supergravity potentials are not easily related\foot{There
are however, some rather interesting results that can be
obtained in this area \dWNWnew.}.  The $SL(2,\IR)$ subgroup
of $SL(8,\IR)$ is in no way special in the four-dimensional
gauged supergravity; all $35$ scalars are on the same footing.
If one turns on scalars in the ${\bf 35}_v$ only, the round $S^7$
upon which the M theory is compactified is deformed to
the ellipsoidal  surface defined by \squish\ with $P=8$.

The foregoing supergravity picture leads to a generalized
class of $F$-theory compactifications.   Recall that M-theory on a $T^2$
is dual to \tobe\  on $S^1$, and that the complex structure
modulus of the $T^2$ is dual to the dilaton and axion, while
the K\"ahler modulus of the $T^2$ is dual to the radius of
the $S^1$ \refs{\PAdual,\JSPow}.  Thus we can identify the M-theory
$T^2$ with the torus of $F$-theory, and this generalizes to elliptically
fibered Calabi-Yau manifolds.  From the discussion of the $S^7$
compactification of M-theory we see a very similar structure:  The dilaton and
axion parameters of the IIB theory have been folded into the
$7$-metric.  To be more precise, combining the $S^7$ with the radial
coordinate,
one has a compactification of M-theory on an $8$-manifold.  The shape of the
metric in two directions of this $8$-manifold is parametrized by the
$SL(2,\IR)$ of the \tobe\  theory, and the scale of this $2$-metric
is dual to the radius of the circle upon which the \tobe\  $D3$-branes
are wrapped.  The $8$-manifold may well be a singular elliptic
fibration, and the full $11$-metric is a warped product, but the
compactification of M-theory on the $8$-manifold, and of the IIB theory
on the $7$-manifold of the radial coordinate and $S^5 \times S^1$,
is a generalization of the usual F-theory and M-theory story.

In this paper we will examine these ideas for the $\none$
supersymmetric flow in which one chiral multiplet is given a mass, and
in the infra-red this field may be ``integrated out'' to leave a
non-trivial conformal fixed point theory.  The holographic version of
this has been extensively studied for the flow of $\nfour$ Yang-Mills
to the $\none$ supersymmetric Leigh-Strassler fixed point
\refs{\RGLMJS \KPW \KLM\FGPWa-\KPNWa,\KPNWc}.  In section 2 we will
review the ten-dimensional solution the \tobe\ supergravity that
corresponds to this flow \KPNWc, and we will rewrite the solution in a
more geometrically transparent manner.  We will show that the solution
of \KPNWc\ is a generalization of the compactification solutions of
\CNPNPW, and by recasting it in this way we see how to create more
solutions in the same class.  In particular, it will lead to a natural
Ansatz for M-theory compactifications.   This Ansatz is further
supported by calculating the  T-dual of the \tobe\ solution
to obtain a solution of IIA supergravity and its lift it to $M$-theory.  

Our primary purpose
here is to examine the analog of the Leigh-Strassler flow in the
scalar-fermion theory in $(2+1)$ dimensions.  The supergravity
critical point was found long ago \refs{\NPWext\NPWpot{--}\HNNW}, and it
has an $SU(3) \times U(1)$ symmetry, and $\ntwo$ supersymmetry ($8$
supersymmetries) in the bulk.  This critical point has several amusing
features and was studied in \HNNW.  More recently this supergravity
solution was studied from the four-dimensional perspective in the
context of RG flows of the scalar-fermion theory
\refs{\Ahnetal,\AhnWoo}.  Indeed, these authors found the $\ntwo$
supersymmetric RG flow solution from the $\neigh$ superconformal
point\foot{Remember that the brane is $(2+1)$-dimensional, and so the
supersymmetry parameters have two components, not four.}.  In section
3 we summarize the relevant results of
\refs{\NPWpot,\NPWext,\HNNW,\Ahnetal}, and then in section 4 we will
use the ideas of section 2 to construct the lift of the
four-dimensional solution to M-theory.   We find that the deformed
geometry of $S^7$ in the lift contains a $\IC \IP^2$ (upon which the
$SU(3)$ acts transitively).  In the same spirit as \CNPNPW, we find
that this $\IC \IP^2$ can be replaced by any Einstein-K\"ahler space,
and in particular by $S^2 \times S^2$.  This leads to a solution with
a  conifold singularity.  More generally, we argue that the solution
of \KPNWc\ and the solutions presented here are examples of a general
class of solutions, and that in $2n+1$  dimensions, there will be 
a solution to Einstein's equations with an $n$-form potential, and
in which there is a $2(n-1)$-dimensional  (real) Einstein-K\"ahler 
submanifold.    

Finally, at the end of section 4 we compute the potential, and
metric on the moduli space of an $M2$-brane probe in our 
solutions.  The results are very similar to those of the $D3$-brane 
probe of the corresponding solution of \tobe\ supergravity\CVJprobe.  
Section 5  contains some final remarks on the structure of our new 
solutions.

\newsec{The geometry of the $\none$ Supersymmetric \tobe\  flow}

\subsec{The flow in ten dimensions}

We first recall some of the results of \FGPWa\ and \KPNWc.
The holographic form of the Leigh-Strassler flow is described
in five-dimensional supergravity by two scalars, denoted
$\alpha$ and $\chi$.  These are respectively dual to the operators:
\eqn\scalops{\eqalign{
\cO_1 ~\equiv~ & \Tr (-X_1^2 - X_2^2 -X_3^2 - X_4^2 + 2\,
X_5^2 + 2\, X_6^2)  \,,\cr
 \cO_3 ~\equiv~ & \Tr (\lambda_3 \lambda_3) \, + \, h.c.  \,\,.}}
The five-dimensional metric is taken to be:
\eqn\fivemet{ds_{1,4}^2 ~=~ dr^2 ~+~ e^{2 A(r)}\big(
 \, \eta_{\mu \nu} \, dx^\mu \, d x^\nu \big) \,.}
The equations describing the flow are then:
\eqn\floweqsred{ {d \alpha \over d r} \eql  {1 \over 6 L}\,
{\del W \over \del \alpha} \ ,\qquad {d \chi \over d r} \eql
{1 \over L}\, {\del W \over \del \chi}  \ , \qquad
{d A \over d r}\eql  - {2 \over 3\,L}\,W   \ .}
where $L$ is the radius of the $AdS_5$, and $W$ is the superpotential:
\eqn\Wpot{W ~=~ {1 \over 4 } \, \rho^4   \,
\big(\cosh(2\chi) - 3 \big)   ~-~  {1 \over 2 \,\rho^2 } \,
 \big( \cosh(2\chi) +1 \big) \, ,}
where $\rho \equiv e^{\alpha}$.

This can then be lifted to a solution of the \tobe\  theory in which
the ten-dimensional metric is given by:
\eqn\warpmetr{ds_{10}^2\eql \Omega^2 \,  ds_{1,4}^2 ~+~ ds_5^2\,,}
where $ds_5^2$ is the metric on the deformed $S^5$ and
$\Omega$ is the warp-factor.  The \tobe\  dilaton and axion
are constant, but there is a non-trivial $B$-field.  The
general metric Ansatz \refs{\KPW,\KPNWc} is given by computing
a non-trivial metric on $\IR^6$ and projecting it onto the unit $S^5$.
Let $x^I, I=1,\dots, 6$ be Cartesian coordinates on $\IR^6$ with $S^5$
defined by $\sum (x^I)^2=1$, then for the flow defined above we have:
\eqn\elliptmet{ds_5^2 \eql L^2 \, {{\rm sech}\chi\over \xi}\,
(dx^IQ^{-1}_{IJ} dx^J)  ~+~ L^2 \, {\sinh\chi\, \tanh\chi\over \xi^3}
\, (x^IJ_{IJ}dx^J)^2\,.}
In this equation $Q$ is a diagonal matrix
with $Q_{11}=\ldots=Q_{44}=\rho^{-2}$ and $Q_{55}=Q_{66}=\rho^{4}$,
$J$ is an antisymmetric matrix with $J_{14}=J_{23}=J_{65}=1$, and
$\xi^2\eql x^IQ_{IJ}x^J$.   The  warp factor is simply
\eqn\elliptwarp{\Omega^2 \eql \xi~\cosh\chi\,. }
We thus see that the metric is a combination of ellipsoidal
squashing of the $S^5$, and a stretching of the Hopf fiber.

Following \KPNWc, we introduce complex coordinates on this
$\IR^6$
\eqn\complcoor{ u^1\eql x^1+i\, x^4\,,\qquad u^2\eql x^2+i\, x^3\,,
\qquad u^3\eql x^5-i\,x^6\,, }
and then reparametrize them using an $SU(2)$ group action:
\eqn\angcoords{ \left(\matrix{u^1\cr u^2\cr}\right)\eql \cos\theta\,
M(\al_1,\al_2,\al_3)\, \left(\matrix{1\cr 0\cr}\right)\,,\qquad
u^3\eql  e^{-i\phi}\,\sin\theta\,,}
where $\al_1,\al_2,\al_3$ are Euler angles.  Associated to $M$ are
the left-invariant $1$-forms: $\sigma_k = \Tr(dM\cdot M^{-1} \, J_k)$.
Explicitly these are:
\eqn\sigdefs{\eqalign{\sigma_1 ~=~ & \cos (\alpha_3)\,  d\alpha_1 +
\sin ( \alpha_1 )\, \sin ( \alpha_3)\, d\alpha_2 \,, \cr
\sigma_2 ~=~ & \sin (\alpha_3)\,  d\alpha_1 -
\sin ( \alpha_1 )\, \cos ( \alpha_3)\, d\alpha_2\,, \cr
\sigma_1 ~=~ & d\alpha_3 + \cos ( \alpha_1 )\,   d\alpha_2  \,.}}
and they satisfy  $d \sigma_i = {1\over 2} \epsilon_{ijk}\,
\sigma_j \wedge \sigma_k$

Finally, define:
\eqn\Xdefn{
X_1(r,\theta)  \eql \cos^2\theta+\rho(r)^6\,\sin^2\theta\,.}
and then the warp-factor, $\Omega$ is given by:
\eqn\Omres{\Omega ~=~ \rho^{-{1 \over 2}}\, (\cosh \chi)^{ 1 \over 2 }\,
X_1^{1\over 4 }\,.}
and the ten-dimensional metric can be diagonalized in terms of the
following frames:

\eqn\theframessp{\eqalign{
e^{\mu+1} &\eql  \rho^{-{1\over 2}}\,(\cosh\chi)^{1\over 2}
\, X_1^{1\over 4} \,e^A\,dx^\mu\,,\qquad \mu=1,\ldots,4\,,\cr
e^5 &\eql \rho^{-{1\over 2}} \,(\cosh\chi)^{1\over 2}\,
X_1^{1\over 4} \,dr\,,\cr
e^6 &\eql L\, \rho^{{3\over 2}}\,   (\cosh\chi)^{{1\over 2}}\,
X_1^{-{3\over 4}}\, \big(\sin^2 \theta \, d\phi + \coeff{1}{2} \,
\cos^2 \theta\, \sigma_3 \big)\,, \cr
e^7 &\eql  L\, \rho^{-{3\over 2}}\,   (\cosh\chi)^{-{1\over 2}}\,
X_1^{{1\over 4}}  \,d\theta\,,\cr
e^8 &\eql L\, \rho^{-{3\over 2}}\,   (\cosh\chi)^{-{1\over 2}}\,
X_1^{1\over 4} \, \sin \theta \, \cos \theta \, \Big( d\phi -
\coeff{1}{2}  \sigma_3 \cr &\qquad \qquad \qquad  ~+~  (1 - \rho^6)  \,
X_1^{-1}\, \big(\sin^2 \theta \, d\phi + \coeff{1}{2} \, \cos^2
\theta\, \sigma_3 \big) \Big)\cr
e^9 &\eql  \coeff{1}{2}\, L \, \rho^{{3\over 2}}\,  (\cosh\chi)^{-{1\over 2}}
\,  X_1^{-{1\over 4}} \, \cos\theta \  \sigma_1 \,,\cr
e^{10} &\eql \coeff{1}{2} \, L\, \rho^{{3\over 2}}\,(\cosh\chi)^{-{1\over 2}}
\,  X_1^{-{1\over 4}} \, \cos\theta \ \sigma_2  \,.\cr }}
It is worth noting that for the round $S^5$, $e^7,\dots,e^{10}$
define the $\IC\IP^2$ base of the Hopf fibration, and the $1$-form:
\eqn\hodfform{ \omega ~\equiv~ \big(\sin^2 \theta \, d\phi + \coeff{1}{2} \,
\cos^2 \theta\, \sigma_3 \big)\,,}
is that of the fiber.

This frame basis is slightly simpler than that of \KPNWc, but the $B$-fields become
much simpler in this system:
\eqn\BAns{\cB ~\equiv~ \coeff{1}{2}\, \cB_{MP} dy^M \wedge d y^P ~=~
\coeff{i}{2}\,\, e^{- i \phi}\,  \sinh  \chi  \,
(e^7 - i \, e^8) \wedge (e^9 - i e^{10}) \,.}

What is apparent here, and was not noted in \KPNWc, is
that the $B$-field is ``doubly-null'' in frames.
That is, it is the wedge of two complex frames whose
norm is zero.  The metric above appears to have an
almost complex structure defined by
\eqn\ACstr{\cJ ~=~ e^5 \wedge e^6 ~+~  e^7  \wedge e^8 ~+~
e^9 \wedge e^{10} \,,}
and if this, or some multiple of it, were integrable, then
the field $\cB$ would carry two holomorphic indices.

We thus see that the foregoing solution is a generalization
of the compactification technique of \refs{\CNPNPW,\LJR}.
This technique was used to create non-trivial compactifications
based upon any Einstein-K\"ahler manifold of complex dimension $n$.
The idea was to introduce a tensor gauge field, $\cA$, of the form
\eqn\PWAns{\cA ~=~ e^{-i k \phi} \, d\zeta_1 \wedge \dots \wedge
d \zeta_n \,,}
where the $\zeta_j$ are complex coordinates on the $n$-fold,
and $\phi$ is the coordinate on the $U(1)$ fibration defined
via the K\"ahler structure.  There is generically no globally
defined holomorphic $(n,0)$-form on the $n$-fold  (unless $c_1 =0$),
but for suitable choice of $k$, \PWAns\ yields a globally defined
form on the total space of the fibration.  It was shown in
\refs{\CNPNPW,\LJR} that one could find interesting anti-de Sitter
compactifications of higher-dimensional supergravities using a background
that involves \PWAns, and in which the metric is that of
the total space of the fibration but with a  ``stretched'' fiber.

One can thus view the solutions of \refs{\KPNWa,\KPNWc} as a generalization
of this class in which the original Einstein-K\"ahler space is also
ellipsoidally deformed, and the space-time metric can be that of
a flow \fivemet\ and not just anti-de Sitter space.  The solutions of
\refs{\CNPNPW,\LJR} were also generically non-supersymmetric, whereas
the whole point of these ellipsoidally squashed solutions is that
they {\it are}  supersymmetric.

As we will see, recasting the solution of \KPNWc\ in the  foregoing manner
will lead to fairly generalizations in M-theory.

\subsec{The T-dual of the  ten-dimensional flow}

We now compactify the $y \equiv x^3$ coordinate in the brane to a
circle of  radius $R_1$ and follow the standard procedure for construction the
T-dual of a string background  \refs{\Buscher,\CVJprim}.    The transformation
of the metric is elementary, and simply replaces $g_{22} \to (g_{22})^{-1}$.
The \tobe\  dilaton is constant, and so the IIA dilaton background is:
\eqn\IIAdil{e^{2 \tilde \phi} ~=~ (g_{22})^{-1} ~=~ {\rho \, e^{-2 A(r)}
\over  X_1 \, \cosh (\chi) } ~=~ e^{-2\, A(r)} \, \Omega^2\,.}
Since there are no components of $\cB_{\mu \nu}$ in the $y$-direction, and
there are no off-diagonal elements of the metric involving $y$, the
$B_{MN}^{NS} \equiv \Re e( \cB_{M N})$ field is unchanged.  The Ramond-Ramond
field, $B_{MN}^{RR}$ similarly has no component in the $y$ direction,
and so its T-dual gives $A^{(3)}_{MN y} = B_{MN}^{RR}$.  The
$A^{(4)}$-field does have a component in the $y$-direction, and its
$T$-dual merely involves dropping this $y$ index.  Thus we get the
following expressions for the IIA backgrounds in terms of the
\tobe\  fields:
\eqn\Tdualtens{A^{(3)}_{MN y} ~=~ \Im m( \cB_{M N} ) \,, \qquad
A^{(3)}_{\mu \nu \rho} ~=~ A^{(4)}_{\mu \nu \rho y}  \,.}
Technically, for a non-zero, but constant axion field, $C^{(0)}$ there is
a corresponding constant IIA background vector field, $A_y ~=~
C^{(0)}$.

It is equally straightforward to lift this solution to M-theory.  Let
$\varphi$ be the extra circle of radius $R_2$, then the corresponding
M-theory metric is:
\eqn\Mmet{ds_{11}^2 ~=~ e^{-{2 \over 3} \tilde \phi}\, ds^2_{IIA} ~+~
e^{{2 \over 3} \tilde \phi} \, (d \varphi + A)^2 \,.}
The radii $R_1$, $R_2$ and the dilaton field thus parametrize
the metric moduli of the torus, $T^2$ defined by $(y, \varphi)$.
The tensor gauge fields lift to eleven dimensions in the obvious manner:
\eqn\Tdualtens{A^{(3)}_{MN \varphi} ~=~ \Re e( \cB_{M N} ) \,, \qquad
A^{(3)}_{MN y} ~=~ \Im m( \cB_{M N} ) \,, \qquad
A^{(3)}_{\mu \nu \rho} ~=~ A^{(4)}_{\mu \nu \rho y}  \,.}

Putting this all together, we arrive at the following metric
in eleven dimensions:
\eqn\metIIA{\eqalign{ds^2_{11} ~=~  & e^{{2\over 3}  A(r)}\,
\Omega^{2 \over 3}\, \big( dr^2 ~+~ e^{2  A(r)} \, \eta_{\mu\nu}
dx^\mu d x^\nu \big) ~+~  L^2\,  e^{{2\over 3}  A(r)} \,
\Omega^{-{16 \over 3}} \, \cosh^4 (\chi)   \, \omega^2  \cr
&~+~ e^{-{4\over 3}  A(r)}\, \Omega^{-{4 \over 3}} \big( d s_2^2 ~+~
\coeff{1}{4} \,L^2\,\rho^2\, e^{2  A(r)} \,  \cos^2 \theta \,
(\sigma_1^2 + \sigma_2^2) \big)\cr
& ~+~ {L^2\, \Omega^{8 \over 3} e^{{2\over 3}  A(r)} \over\rho^{2} \,
\cosh^2 (\chi)}\, \big( d\theta^2 + \sin^2 \theta\, \cos^2 \theta\,
\big(d\phi - \coeff{1}{2} \, \sigma_3 +(1 -\rho^6)\,X_1^{-1}\,
\omega \big)^2 \big)  \,, } }
where $ds_2^2$ is the metric on the flat torus, and
we now have $\mu, \nu, \dots =0,1,2$.

The tensor gauge field background becomes:
\eqn\Mthtens{A^{(3)}_{\mu \nu \rho} ~=~ A^{(4)}_{\mu \nu \rho y}\,,
\qquad  A^{(3)}_{MN z} ~=~   \cB_{M N}  \,,}
where we have introduced the natural complex coordinate
$z = \varphi +  (C_0 + i e^{-\Phi}) y$ on the torus $T^2$,
and we have been more careful in incorporating the effect of the
non-trivial flat metric on $T^2$.

The basic structure of the $F$-theory compactification is
now more evident.  By virtue of the warp-factors, the $T^2$ fiber
is most naturally paired with the $S^2$ upon which the
$SU(2)$ isometry acts.  The internal tensor gauge field,
$A^{(3)}$, is obtained by adding the holomorphic index of
the $T^2$  to the ``$(2,0)$'' structure of the field, $\cB_{MN}$.
It is precisely this structure that we generalize in subsequent
sections.

One should, of course, remember that the foregoing solution
represents a uniform distribution of $M2$ branes spread over the
$T^2$.  This solution is a function of the radial coordinate, $r$,
of $\IR^6$, and not $\IR^8$.  In subsequent sections we will
be looking for $M2$-brane solutions that are localized in $\IR^8$,
and the solution above should be some kind of generalized ``Coulomb
branch'' of the more localized brane distributions.  Based on our
experience of other Coulomb branch flows, the results above suggest
that the more localized $M2$-brane solutions should have a similar
natural complex structure to the metric and to the $B$-field background.
We will indeed find that this is the case.

\newsec{The holographic RG flow in four dimensions}

The analogue of the LS flow in four dimensions is the $\ntwo$
supersymmetric RG flow in $\neigh$, four-dimensional  gauged
supergravity  constructed in \refs{\Ahnetal,\AhnWoo}. 
The two scalar fields, $\lambda$ and
$\lambda'$, in this flow parametrize an $SU(3)\times U(1)$ invariant
subspace of the full scalar manifold $E_{7(7)}/SU(8)$. The explicit
dependence of the scalar 56-bein,
\eqn\scalviel{
{\cal V}(\lambda,\lambda')\eql \left(\matrix{u_{ij}{}^{ab} &
v_{ijcd}\cr v^{klab}&u^{kl}{}_{cd}\cr}\right)\,, }
on  $\lambda$ and $\lambda'$ fields has been obtained
in \refs{\NPWext\NPWpot\HNNW\Ahnetal-\AhnWoo}. We recall that
all indices $i,j$ and $a,b$ in \scalviel\ run from 1 to 8 and
correspond to the realization of $E_{7(7)}$ in the $SU(8)$
basis\foot{The $SU(8)$ basis corresponds to the use of
${\bf 8}_s$-indices of $SO(8)$. In this basis, the scalars
(${\bf 35}_v$) and pseudoscalars (${\bf 35}_c$) are represented by
self-dual and ati-self-dual $4$-forms.   To compute the metric and
to compare operators in the dual SCFT, it is more convenient to
use the $SL(8,\IR)$ basis.  This is a triality  rotation of the
$SU(8)$ basis in which the ${\bf 8}_s$-indices are converted to
${\bf 8}_v$-indices using gamma matrices.},
$(z_{ab},\bar z^{ab})$. We refer the reader to either \HNNW, or
the Appendix of \Ahnetal\ for the explicit results that we will use here.

The structure of the scalar sector of the $\neigh$ supergravity is
encoded in the $SU(8)$ T-tensor \dWNic:
\eqn\ttensor{
T_l{}^{kij}\eql \left(u^{ij}{}_{ab}+v^{ijab}\right)
\left(u_{lm}{}^{bc}u^{km}{}_{ca}-v_{lmca}v^{kmca}\right)\,. }
In particular, the superpotential, $W$, for the flow is found as one
of the eigenvalues of the symmetric tensor
\eqn\theatens{
A_{1}{}^{ij}~\equiv~ -{4\over 21}T_m{}^{ijm}\,,\qquad W\eql
A_1{}^{77}\eql A_1{}^{88}. }

To preserve the analogy with the five-dimensional flow,  we
introduce new fields
\eqn\newfields{
\alpha\eql{\lambda\over 4\sqrt2} \,,\qquad
\chi \eql {\lambda'\over\sqrt2}\,,}
and define $\rho\eql e^\alpha$. Indeed, upon rotation to the
$SL(8,\IR)$ basis $(x^{IJ},y_{IJ})$ \ECBJ,
\foot{An explicit realization of the $SO(8)$ matrices
$\Gamma_{IJ}$ we use here can be obtained as follows: Starting with
the $SO(7)$ gamma matrices in Appendix C.1 of \FGPWa\ we define
$$
\Gamma_I\eql -i \Gamma_I^{\rm FGPW}\,,\qquad
\Gamma_7\eql -i \Gamma_0^{\rm FGPW}\,,\qquad
I=1,\ldots,6\,.
$$
Then they are all real, antisymmetric and square to $-1$ and we have
$$
\Gamma_{IJ}\eql {1\over 2}\,[\Gamma_I,\Gamma_J]\,,\qquad
\Gamma_{I8}\eql -\Gamma_I\,.
$$
Then  $I,J$ are set to to run from $1$ to $8$.
}
\eqn\baschange{
Z_{ab}\eql {1\over 4}{x^{IJ}+i
y_{IJ}\over\sqrt2}\,(\Gamma_{IJ})_{ab}\,, }
the generator, $Q$, corresponding to the $\alpha$ field is diagonal in
$SL(8,\IR)$
\eqn\algen{
Q\eql \left(\matrix {t{}^{[ij]}{}_{[ij]} & 0\cr
0 & t{}_{[kl]}{}^{[kl]}\cr}\right)\,,
}
where
\eqn\thet{
t{}^{[ij]}{}_{[ij]}\eql q^i+q^j\,,
\qquad
t{}_{[ij]}{}^{[ij]}\eql q_i+q_j\,,\qquad q_i\eql - q^i\,,}
and
\eqn\theq{
(q^i)\eql {1\over 2}\,(-1,+3,+3,-1,-1,-1,-1,-1)\,,}
which is the counterpart of the similar result in five dimensions.
In subsequent sections we will make an $SO(8)$ rotation of this
so as to place the two $+3$ eigenvalues in the last two
entries of $q^i$.

In terms of new fields the superpotential is  \Ahnetal:
\eqn\superpot{
W(\alpha,\chi)\eql {1\over 8} {\rho^6}\,
(\cosh(2\chi)-3) ~-~ {3\over 8\,\rho^2}\,(\cosh(2\chi) + 1)\,,}
and the supergravity potential is then given by:
\eqn\sugrapot{
\cP(\alpha,\chi) \eql {4\over  L^2}\, \Big[{1 \over 6}\, \Big(
{\del W \over \del \alpha} \Big)^2 ~+~  \Big({\del W \over \del
\chi} \Big)^2 - 3\, W^2\Big]\,.}

The study of supersymmetric flows closely parallels the discussion
in \FGPWa.  One considers a metric of the form \fivemet, but now with
$\mu,\nu =0,1,2$.  One then finds that the supersymmetric flow equations
are given by \Ahnetal:
\eqn\flowfora{ A'(r)\eql -{2\over L}\   W\,,}
and
\eqn\flowfields{ \eqalign{ 
{d\rho\over dr}&\eql  {1\over 8\, L}\,
{(\cosh(2\chi) + 1)~+~  (\cosh(2\chi)-3) \rho^8 \over \rho}\,,\cr
{d\chi\over dr}&\eql {1 \over  2\,L}\,{(\rho^8 - 3)
\sinh(2\chi) \over \rho^2}\,.\cr}}
{}From this it is evident that there is a supersymmetric critical point
at   $\rho = 3^{1/8}$, $\cosh(2\chi) =2$.  At this point we have
$W  = -{1 \over 2} 3^{3/4}$.   This is the $\ntwo$ supersymmetric
critical point found in \NPWext, and studied in \HNNW.  The flow that
we are primarily interested in is the one that starts at the
$\neigh$ point, and finishes at the non-trivial $\ntwo$ supersymmetric
point.
 
It is a useful exercise to determine the operators which are dual to
the supergravity fields $\rho$ and $\chi$. In~\HNNW, the scalar
expectation value, $\rho$, was given by a self-dual form written in
the $SU(8)$ basis. An $SO(8)$-triality rotation of this self-dual form 
to the $SL(8,\IR)$ basis results in an $8 \times 8$, symmetric, traceless 
matrix  from which one can read off the dual operator:
$\cO^{(\rho)}=\cO^{77} +\cO^{88}$, where $\cO^{IJ}$ was defined
in~\bilinears. The pseudoscalar $\chi$ is given by an anti-self-dual
form which, after a triality rotation is found to
be dual to $\cP^{(\chi)}=\cP^{33} -\cP^{44}$.  Despite the apparently
disparate indices, these two operators do indeed lie in the same
supermultiplet on the brane.

\newsec{Generalizing the supersymmetric flow}

Our goal is ultimately to construct the lift to $M$-theory of the
solution  described in section 3.  We will, however, proceed rather
more generally and start by abstracting some ideas
from the flow described in section 2.  We will thus be first led
to an Ansatz for $S^{2n+1}$, and we will then implement it on an
$S^7$ in $M$-theory.

\subsec{Stretching and squashing spheres}

The metric we ultimately want is an ellipsoidally squashed
sphere with a stretched Hopf fiber.  To set our notation, and
explain this terminology, we begin with a brief review of
the Hopf fibration and its stretching.

Introduce Cartesian coordinates, $x^I$, $I=1,\ldots,2n+2$, on
$\IR^{2n+2}$ and think of $S^{2n+1}$ as defined by
the surface $\sum_I(x^I)^2=1$.  Now define complex coordinates:
\eqn\complCoor{
z^1\eql x^1+i x^2\,,\quad \ldots\quad \,,\, z^{n+1}\eql
x^{2n+1}+ix^{2n+2}\,, }
and an associated K\"ahler form, $J_{IJ}$, with:
\eqn\theJJ{
J_{12}\eql J_{34}\eql\ldots \eql J_{2n+1\,2n+2}\eql 1\,. }
Introduce projective coordinates, $\zeta_j$, and the Hopf
fiber angle, $\psi$, via:
\eqn\ccor{
z^i\eql \zeta^i z^{n+1}\,,\quad i=1,\ldots,n\,;\qquad z^{n+1}
\eql (1+\zeta^i\bar\zeta_i)^{-1/2}\,e^{i\psi}\,.}
In these coordinates the metric on $S^{2n+1}$ becomes:
\eqn\rndmet{ds^2 ~=~  \big(d\psi ~+~ A_{(n)}\big)^2~+~
 ds^2_{FS(n)}\,,}
where $ ds^2_{FS(n)}$ is the Fubini-Study metric on $\IC\IP^n$,
and $A_{(n)}$ is the potential for the K\"ahler form
on $\IC\IP^n$.  More explicitly, we have:
\eqn\FSmetric{
ds_{FS(n)}^2\eql {d\zeta^i d\bar\zeta_i\over 1+\zeta^i\bar\zeta_i}
-  {(\zeta^i d\bar\zeta_i)(\bar\zeta_j d\zeta^j)\over
(1+\zeta^i\bar\zeta_i)^2}\,, }
and
\eqn\Apot{
A_{(n)}\eql -{i\over 2}{\bar\zeta_id\zeta^i-\zeta^id\bar\zeta_i\over
1+\zeta^i\bar\zeta_i}\,,}
One may also verify the following rather useful identities:
\eqn\fsmetr{ ds_{FS(n)}^2 \eql (dx)^2-(xJdx)^2 \,, \qquad
 d\psi~+~ A_{(n)} \eql xJdx \,. }

The metric on the stretched sphere  \CNPNPW\ is thus given by:
\eqn\Stretch{ \eqalign{
ds^2(\chi)&\eql \left((dx)^2-(xJdx)^2 \right)+\cosh^2(\chi)(xJdx)^2\cr
&\eql ds^2_{FS(n)}+\cosh^2(\chi)\,\left(d\psi+A_{(n)}\right)^2\,,\cr}}
The parameter, $\chi$, represents the stretching factor, with
$\chi=0$ corresponding to the round sphere. The isometry of $S^{2n+1}$
is, of course, $SO(2n+2)$, and stretching the Hopf fiber breaks this
to $U(n+1)$.

The metrics we wish to construct not only have this stretched
fiber, but are also ellipsoidally squashed, and this further
reduces the isometry  to $SU(n)\times U(1)^2$.
The ellipsoidal squashing is done by following the construction of
\elliptmet, and by once again making a non-trivial metric in
$\IR^{2n+2}$ and then projecting it onto the unit $S^{2n+1}$
\BdWHNNW.  Let  $Q$ be a diagonal matrix
\eqn\theQQ{
Q\eql {\rm diag}(\rho^{-2},\ldots,\rho^{-2},\rho^{2n},\rho^{2n})\,,}
The metric on the deformed $\IR^{2n+2}$ is then given by:
\eqn\theMetr{
ds^2(\rho,\chi)\eql  dx^IQ^{-1}_{IJ}dx^J + {\sinh^2(\chi)\over \xi^2}
(x^I J_{IJ} dx^J)^2\,, }
where $\xi^2\eql x^IQ_{IJ}x^J$.  The metric \elliptmet\ is simply
$ L^2 (\xi  \cosh \chi)^{-1} ds^2(\rho,\chi)$ with $n=2$.  This class
of metrics, for $\chi =0$, were also obtained in the study of
consistent truncations in \CveticTB.

To write the metric \theMetr\ in terms of intrinsic coordinates on 
$S^{2n+1}$, we split the coordinates according to the eigenvalues of $Q$ 
by setting:
\eqn\splitCoord{ x^i\eql \cos\mu\, u^i\,,\quad i=1,\ldots,2n\,;\qquad
x^{2n+1,2}\eql \sin\mu\, v^{1,2}\,, }
where $u^i$ parametrize a unit $S^{2n-1}$ and $v^{1,2}$ a unit circle.
Using \complCoor\ and \ccor, we verify that
\eqn\idHopf{ vJdv\eql d\psi\,, \qquad (uJdu)\eql d\psi ~+~
{ A_{(n)}\over \cos^2\mu}\,.   }
It is now straightforward to verify that the metric \theMetr\ can be
written in the following diagonal  form
\eqn\finalmetric{ \eqalign{
ds^2(\rho,\chi) & \eql
\big(\, \rho^{-4}\xi^2\, d\mu^2+
 \rho^2\,\cos^2\mu\ ds_{FS(n-1)}^2
+  \xi^{-2}\,\omega^2 \,\big) ~+~
\cosh^2\chi\, \xi^{-2}\, \big(d\psi+A_{(n)}\big)^2 \,.\cr} }
where
$$
\omega\eql {1\over 2}(\rho^4-\rho^{-4})\,\sin(2\mu)\,d\psi
+\rho^4\tan\mu\,A_{(n)}\,.
$$
Comparing \finalmetric\ with \Stretch, we see that the non-trivial
squashing deforms the metric on the $\IC\IP^n$ base and rescales the
Hopf fiber, but preserves the $\IC\IP^{n-1}$, whose symmetry
group is $SU(n)$.   It should be remembered that $A_{(n)}$ is
the vector potential for the K\"ahler structure on $\IC\IP^n$,
and if one decomposes it into components on  $\IC\IP^{n-1}$, then
another angular coordinate, $\tilde \psi$,  emerges in $A_{(n)}$.
There are thus two $U(1)$ symmetries, namely  rotations in $\psi$
and $\tilde \psi$.

This leads to a rather natural Ansatz for an $n$-form potential
on the deformed $S^7$:
\eqn\nform{C^{(n)} ~=~ F(\chi, \rho) \, e^{i(\kappa_1 \psi +
\kappa_2 \tilde \psi)} \, (e_1 - i e_2) \wedge \ldots \ldots
\wedge (e_{2n-1} - i e_{2n}) \,,}
In this equation the $e_j$ are an orthonormal frame such that
$(e_{2j-1} - i e_{2j})$, $j=1,\dots,n-1$ are holomorphic frames on
$\IC\IP^{n-1}$, $F$ is generically an arbitrary function of
$\rho$ and $\chi$ (but we will specify it more completely
below), and $\kappa_a$ are constants to be determined by
requiring that $C^{(n)}$ be globally defined, and further fixed
by the determining the  unbroken $U(1)$ symmetry.

\subsec{ The $\ntwo$ supersymmetric flow in $M$-theory}

Given the known results for consistent truncation of gauged
$\neigh$ supergravity in four dimensions \refs{\ConTrunc,\BdWHNNW},
we can obtain the metric of deformed $7$-sphere compactification
of $M$-theory rather directly.  To be explicit, we have:
\eqn\elevenmet{ds_{11}^2 ~=~ \Delta^{-1} \big( dr^2 + e^{2 A(r)}
(\eta_{\mu \nu} dx^\mu dx^\nu) \big) ~+~  ds_7^2 \,,}
where $\mu, \nu = 0,1,2$.  The inverse metric on $S^7$ is given by
\BdWHNNW:
\eqn\EmetAns{ \Delta^{-1}\,g^{pq} \eql
(K_{MN})^p \,(\Gamma_{MN})^{ab}(u_{ij}{}^{ab}+v_{ijab})
(u^{ij}{}_{cd}+v^{ijcd}) (\Gamma_{PQ})^{cd}\, (K_{PQ})^q\,,}
where $K_{MN}=x^M\partial_N-x^N\partial_M$, and $x^I$ are coordinates
in $R^8$.  As usual, the warp factor, $\Delta$, is defined by:
\eqn\Deltadefn{\Delta \,\equiv\,  \sqrt{{\rm det}( g_{mp}\,
\gop^{pq})} \,,}
where the inverse  metric, $\displaystyle {\gop^{pq}}$, is that
of the ``round''  $S^7$.  One can compute $\Delta$ by taking the
determinant of both sides of \EmetAns.  In \EmetAns\ we have also
inserted $\Gamma$-matrices so as to triality rotate the $SO(8)$
Killing vectors.  This has the effect of changing the $SU(8)$
indices  on $u,v$ to those the $SL(8,\IR)$ basis of $E_{7(7)}$, and
it in this basis that the ellipsoidal
squashing is more directly visible.

Using this formula,  we find the following form for the metric 
on $S^7$:
\eqn\Sseven{ds_7^2 ~=~ \Delta^{1 \over2}\, ds^2(\rho,\chi) \,,}
where $ds^2(\rho,\chi)$ is given by \theMetr, $\xi^2 =x^IQ_{IJ}x^J$,
and
\eqn\Deltres{\Delta ~=~ \big(\xi\, \cosh \chi\big)^{-{4\over 3}} \,.}
In particular, we recover a metric that is conformally related
to one of the metrics described in the previous subsection.

We now introduce a spherical parametrization of this metric in
a manner closely analogous to \KPNWc.  The coordinates $u^i$
and $v^a$ of \splitCoord\ are replaced according to:
\eqn\spherparam{\eqalign{
u^1+ i\, u^2& \eql \sin\theta\, \cos(\coeff{1}{2}\, \alpha_1) \,
   e^{{i \over 2} (\alpha_2 + \alpha_3)} \, e^{i( \phi + \psi)}  \,,\cr
u^3+ i\, u^4& \eql \sin\theta \,  \sin(\coeff{1}{2}\, \alpha_1) \,
    e^{-{i \over 2} (\alpha_2 - \alpha_3)} \, e^{i( \phi + \psi)}   \,,\cr
u^5+ i\, u^6 & \eql  \cos\theta\,  e^{i(\phi + \psi) } \,,\cr
v^1 + i\, v^2 & \eql  e^{i \psi }\,.  }}
The coordinate, $\psi$, is that of the Hopf fiber  on
$S^7$, while $\psi+\phi$ is the Hopf fiber coordinate of the $S^5$
defined by the $u^i$.

The left-invariant $1$-forms are given by \sigdefs, and one can
easily rewrite the metric on $\IC\IP^2$ in terms of them:
\eqn\CPtwo{ds_{FS(2)}^2 \eql   d\theta^2 ~+~ \coeff{1}{4}\,
\sin^2\theta \, \big( \sigma_1^2+\sigma_2^2 + \cos^2\theta\, \sigma_3^2
\big)\,.}
Similarly, the metric in $\IC\IP^3$ may be written:
\eqn\CPthree{ds_{FS(3)}^2 \eql d\mu^2 ~+~ \cos^2\mu \,\big(d_{FS(2)}^2 ~+~
\sin^2\mu\, (d\phi ~+~ \coeff{1}{2}  \sin^2\theta\, \sigma_3)^2\big) \,.}
Using these coordinates, we find the following set of frames for the
the eleven-dimensional metric \elevenmet:
\eqn\frames{ \eqalign{
e^{\mu+1}& \eql e^A\, (\cosh\chi)^{2/3}\,{X^{1/3}\over \rho^{2/3}}\,
dx^\mu\,,\qquad \mu=0,1,2\,,\cr
e^4&\eql (\cosh\chi)^{2/3}\,{X^{1/3}\over \rho^{2/3}}\,dr\,,\cr
e^5&\eql a\, (\sech\chi)^{1/3}\,{X^{1/3}\over \rho^{8/3}}\, d\mu\,,\cr
e^6&\eql a \, (\sech\chi)^{1/3}\,{\rho^{4/3}\over X^{1/6}}\,\cos\mu\,d\theta
\,,\cr
e^7&\eql {a\over 2}\,  (\sech\chi)^{1/3}\,{\rho^{4/3}\over X^{1/6}}\,\cos\mu\,
\sin\theta\,\sigma_1\,,\cr
e^8&\eql {a\over 2} \, (\sech\chi)^{1/3}\,{\rho^{4/3}\over X^{1/6}}\,\cos\mu\,
\sin\theta\,\sigma_2\,,\cr
e^9&\eql {a\over 4} \, (\sech\chi)^{1/3}\,{\rho^{4/3}\over X^{1/6}}\,\cos\mu\,
\sin(2\theta)\,\sigma_3\,,\cr
e^{10}&\eql {a\over 2}
 \, (\sech\chi)^{1/3}\,{\rho^{16/3}\over X^{2/3}}\,\sin(2\mu)\,
\big[\,(1-{1\over \rho^8})\,d\psi+
(d\phi+{1\over 2}\sin^2\theta \sigma_3)\,\big]\,,\cr
e^{11}&\eql {a}\,
(\cosh\chi)^{2/3}\,{\rho^{4/3}\over X^{2/3}}  \,
\big[\, d\psi+\cos^2\mu\,(d\phi+{1\over 2}\sin^2\theta \sigma_3)\,
\big]\,,\cr}}
where the constant, $a$, will be fixed momentarily, and
$$
X(r,\mu) ~\equiv~ \cos^2\mu+\rho(r)^8\sin^2\mu\,.
$$

In computing the Ricci tensor we use the equations of motion
\flowfora\ and \flowfields.  For $\rho=1$ and $\chi=0$ we must recover
the $AdS_4 \times S^7$ solution in which one has:
\eqn\RndRicci{ R_{AB}\eql {6\over L^2}\, {\rm diag}
\left(\matrix{2, &-2, &-2, &-2, &1, &\ldots &, &1}\right)\,,}
where $A,B$ are frame indices.  This fixes the constant, $a$, according to:
\eqn\sphrad{a\eql\,L\,, }
that is the radii of the $AdS_4$ and $S^7$ are $L/2$ and $L$, respectively.
We also find that the general Ricci tensor has only two non-vanishing 
off-diagonal components: $R_{45}$ and $R_{10\,11}$.  This tensor
also satisfies obvious symmetries due to the Poincar\'e and
$SU(3)$ invariance but it also satisfies a non-trivial
identity parallel to the one found in \KPNWc.  Thus we find
in general that:
\eqn\ricciid{
R_{11}\eql -R_{22}\eql -R_{33}\eql 2R_{66}\eql 2 R_{77}\eql 2 R_{88}
\eql 2R_{99}\,,}
where all the indices are frame indices.

For the antisymmetric field $F^{(4)}$ we take an ansatz similar to that
of \KPNWc, and motivated by \PWAns.  First note that for $\chi=0$
and $\rho=1$ the internal metric of \frames\ contains a
$\IC\IP^3$ factor, and that the K\"ahler form of this, when
written in terms of frames, is
\eqn\kform{
{\cal J}\eql {1\over 2} L^2\, dA_{(3)}\eql e^5\wedge e^{10}+e^6\wedge
e^9+e^7\wedge e^8\,,}
which implies that the natural basis of the holomorphic 1-forms 
consists of
\eqn\holforms{
e^5-ie^{10}\,,\qquad e^6-ie^9\,, \qquad e^7-ie^8\,.}
We thus take the internal part of $A^{(3)}$ to be the real part of:
$$
C^{(3)}  ~\equiv~ c \, \sinh\chi\,e^{ i( \kappa_1 \psi+
\kappa_2\phi)}\, (e^5-ie^{10})\wedge ( e^6-ie^9)\wedge ( e^7-ie^8) \,,
$$
where $c$, $\kappa_1$ and $\kappa_2$ are some real constants.
This is of the form \nform, and the arbitrary function $F(\rho,
\chi)$ is now fixed by the proper choice of frames, and  
through comparison with \BAns.

As is implied by \Mthtens, the tensor $A^{(3)}$ also has a
space-time part that is very similar to the Ansatz for the
$A^{(4)}$-tensor in the \tobe\  theory.  We therefore take:
\eqn\Athree{ A^{(3)} \eql \widetilde W(r,\mu)\, e^{3 A(r)} \, dx^0
\wedge dx^1 \wedge dx^2  ~+~  (C^{(3)} ~+~ (C^{(3)})^*)\,,}
where  $\widetilde W(r,\mu)$ is a ``geometric superpotential''
to be determined.

The equations of motion (in the conventions of \CNPNPW) are:
\eqn\eqnmot{ R_{MN} ~+~ R \, g_{MN}  \eql\coeff{1}{3}\, F^{(4)}_{MPQR}\,
F^{(4)}_N{}^{PQR}\,,\qquad d*F^{(4)}\eql F^{(4)}\wedge F^{(4)}\,, }
where $*$ is defined using $\epsilon^{1\cdots 11}\eql 1$.

Starting with the Einstein equations, one finds that the
right-hand side has generically non-vanishing off-diagonal
terms whereas the corresponding components of the Ricci tensor
vanish.   These off-diagonal components can be made to vanish by setting:
\eqn\thekappas{
\kappa_1\eql -4\,,\qquad \kappa_2\eql- 3\,.
}
The (10,11)-component determines $c$ (up to a sign), and so
we have
\eqn\thec{
c\eql {1\over 4}\,.
}
The (4,4), (4,5) and (5,5) components may be used to determine
$\widetilde W$ (again, up to a sign), and we find
\eqn\Wtilde{\widetilde W(r,\mu) \eql  {1 \over 4\, \rho^2 }\,
\big[\, (\cosh(2\chi)+1)\,\cos^2\mu ~-~ \rho^8\,(\cosh(2\chi)-3)\,
\sin^2\mu\, \big]\,.}
With these values for the constants and using \Wtilde, we find that
all of the equations of motion of $M$-theory are indeed satisfied.

\subsec{Brane probes}

It is relatively straightforward to  perform  a brane probe
calculation  of the  supergravity solution presented above, and the
results are directly parallel those of \refs{\CVJprobe,\AKNW}.   
This is perhaps not
surprising since dimensional reduction and T-dualization of the
\tobe\  solution effectively adds one more complex scalar, and
thereby extends the Coulomb moduli space.

The $M2$-brane calculation is very similar to the $D$-brane
probe calculation.  One starts with an action:
\eqn\BPact{S ~=~  \int \, d^3 \sigma\, \Big[  \sqrt{
- det(\tilde g)}  ~+~ \coeff{1}{3}\, \tilde A^{(3)}\Big] \,.}
where $\tilde g$ and $\tilde A^{(3)}$ denote the pull-back of the
metric and the $3$-form onto the membrane.  The  normalization of the
$A^{(3)}$-term in \BPact\ is twice the usual normalization since
this is the normalization that we have used in the eleven-dimensional
equations of motion.   As usual, we consider
a probe that is parallel to the source membranes, and assume that
it is traveling at a small velocity transverse to its world-volume.
This calculation produces a potential, $V$, and a kinetic term for the
brane probe.  If the potential vanishes, then the kinetic term
provides us with a metric on the corresponding moduli space, and this
metric has the form $h_{ab} = \delta^{1/2} \,(g_{00})^{-1} g_{ab}$, 
where $g_{MN}$  is the eleven-dimensional metric,  $a,b$ index coordinates
transverse to the $M2$-branes, and $\delta$ is the determinant
of the projection of $g_{MN}$ parallel to the brane.

We find the following expression for the potential:
\eqn\probpot{ V ~=~   e^{3 A(r)} \, \big(\Delta^{-{3 \over 2}} ~-~ 2\,
\widetilde W \big) ~=~ 2\, e^{3\, A(r)}  \, \rho^6\,  \sinh^2  \chi
\, \sin^2 \mu \,,}
which is very similar to that found in \CVJprobe.

This potential vanishes for $\mu =0$, and on this subspace we  
have the following metric on the $6$-dimensional moduli space 
transverse to the branes:
\eqn\metonmod{ ds^2 ~=~  \Delta^{-{3 \over 2}}\,  e^{ A} \,
\big[  dr^2 ~+~ L^2\, \rho^4 \, \sech^2 \chi \, ds^2_{FS(2)} ~+~
 L^2\, \rho^4 \, (d\psi + d\phi + \coeff{1}{2} \sin^2 \theta \,
\sigma_3)^2 \big]  \,.}
As one approaches the critical point one can introduce a 
new radial coordinate, $u \sim e^{{1 \over 2} A(r)}$ to obtain the
following asymptotic form of the metric:
\eqn\asympmetonmod{ ds^2 ~\sim~   du^2 ~+~ {3 \over 2}\, u^2 \,
ds^2_{FS(2)} ~+~{9 \over 4}\, u^2 \, (d\psi + d\phi + \coeff{1}{2} 
\sin^2 \theta \, \sigma_3)^2  \,.}
This form is very similar of that found in \CVJprobe:  In the 
latter $D3$-brane probe calculation, the asymptotic metric had a 
similar conical singularity at $u=0$, with the $\IC \IP^2$ replaced
by $S^2$, and the ``stretching factors'' ${3 \over 2}$ and
${9 \over 4}=\big({3 \over 2}\big)^2$ replaced by
 ${4 \over 3}$ and ${16 \over 9}=\big({4 \over 3}\big)^2$.
The meaning of these conical singularities in moduli space 
has yet to be adequately explained, but they may indicate that
the supergravity coordinates are not appropriate to the
correct description of the moduli space.  We understand that
a forthcoming paper \CVJnew\ will greatly elucidate this issue.

\subsec{Generalizations and conjectures}

The construction of solutions of \CNPNPW\ was done in two
steps:  the first was to obtain a solution on a deformed $S^7$,
and the second step was to generalize the result to the
canonical $U(1)$ bundle over an arbitrary Einstein-K\"ahler manifold.
One can obviously try to do the same thing here.
It is not clear whether one could replace the $\IC\IP^3$ in our
construction by an arbitrary Einstein-K\"ahler $3$-fold.  On the
other hand it does seem very plausible that one could replace the
$\IC\IP^2$ parametrized by $\theta, \alpha_j$ by an arbitrary
Einstein-K\"ahler $2$-fold.  This is because this space is homogeneous
and presumably the eleven-dimensional Ricci tensor only depends upon
whether the $2$-fold is  Einstein, and if there  is a canonical
$U(1)$ bundle whose total space can also be made into an Einstein space.

Einstein metrics on K\"ahler manifolds have been fairly extensively
studied.  In particular, for $2$-folds there are the obvious ones:
$\IC \IP^2$ and $S^2 \times S^2$, and also the less obvious:
there are Einstein-K\"ahler metrics on the del Pezzo surfaces, $P_k$,
for $k\ge 3$ (for further discussion of this  see, for
example, \refs{\Kehagias\AFHS{--}\DPEins}).  The total spaces
of the canonical $U(1)$ bundles over these spaces can be made into
an Einstein-Sasaki manifold, and have been used for compactifications of
supergravity without fluxes.  The trivial case is the sphere,
$S^{2n+1}$, as a $U(1)$  bundle over $\IC\IP^n$.  A less trivial
example in the $T^{1,1}$ space used in \LJR\ to compactify
\tobe\  supergravity.  This space is, of course, a $U(1)$ bundle over
$S^2 \times S^2$, and even more significantly, it preserves some
of the supersymmetry of the \tobe\  theory.
One can include fluxes in this compactification in much the same manner
as \CNPNPW.  Indeed, \nform\ generalizes in an obvious manner.
Moreover, if the Einstein metric on the $U(1)$ fibration leads
to a supersymmetric compactification without fluxes, then one might hope
that inclusion of a flux might preserve some supersymmetry.

While we have not tested all of these ideas, we have at least
considered the metric defined from \frames, but with $\IC \IP^2$
replaced by $S^2 \times S^2$.  To be more precise, we consider
frames of the form \frames, but with:
\eqn\newframes{ \eqalign{
e^6&\eql a\, b \, (\sech\chi)^{1/3}\,{\rho^{4/3}\over 
X^{1/6}}\,\cos\mu\,d\theta_1
\,,\cr
e^7&\eql  a\, b \,  (\sech\chi)^{1/3}\,{\rho^{4/3}\over X^{1/6}}\,\cos\mu\,
\sin\theta_1 \,d \varphi_1\,,\cr
e^8&\eql  a\, b\, (\sech\chi)^{1/3}\,{\rho^{4/3}\over X^{1/6}}\,\cos\mu\,
d\theta_2 \,,\cr
e^9&\eql  a\, b \, (\sech\chi)^{1/3}\,{\rho^{4/3}\over X^{1/6}}\,\cos\mu\,
\sin\theta_2 \,d \varphi_2\,,\cr
e^{10}&\eql {a\over 2}
 \, (\sech\chi)^{1/3}\,{\rho^{16/3}\over X^{2/3}}\,\sin(2\mu)\,
\big[-\rho^{-8} \,d\psi+ e\,
A\,\big]\,,\cr
e^{11}&\eql {a}\,
(\cosh\chi)^{2/3}\,{\rho^{4/3}\over X^{2/3}}  \,
\big[\,\sin^2 \mu \, d\psi + e \cos^2\mu \,A \big]\,.\cr}}
where
\eqn\thestwoA{
A\eql (d\psi+ d\phi+ \cos\theta_1
\, d\varphi_1 + \cos\theta_2 \,  d\varphi_2)
}
The frames, $e^6,\dots,e^9$ are proportional to those of
$S^2 \times S^2$.  Observe that the Hopf fiber connections
$\cos\theta_j \, d\varphi_j$ appear in $e^{10}$ and $e^{11}$
with equal weight, and thus $S^2 \times S^2$ along with the
coordinate $(\phi + \psi)$ map out the $T^{1,1}$ space.
The value of $a, b$ and $e$ can be fixed so as to make the Ricci
tensor have the form  \RndRicci\ for $\chi=0$ and $\rho=1$.
Indeed, this fixes these constants to:
\eqn\theconst{
a \eql    L  \,, \qquad  b \eql {1 \over \sqrt{6}}
\,, \qquad  e \eql {1 \over 3}   \,.
}
These values of $b$ and $e$ are precisely those that give the
$T^{1,1}$ space.

First, we check that, when written with frame indices, the Ricci
tensor of this new metric is {\it identical} to that defined by
\frames\ provided that one uses exactly the same equations of motion
for $\chi$ and $\rho$.  This supports the conjecture that the Ricci
tensor is indeed independent of the particular Einstein-K\"ahler
$2$-fold.  Secondly, one can obviously generalize the Ansatz for
$A^{(3)}$.  In fact, we find that a minor modification of \Athree,
where the geometric superpotential, $\widetilde W(r,\mu)$, is kept the
same while 
\eqn\neCthree{
C^{(3)}\eql - {1\over 4}\, \sinh\chi\,e^{-\phi-2\psi}\,
(e^5-ie^{10})\wedge (e^6+ie^7)\wedge (e^8+ie^9)\,, }
yields a solution to the equations of motion.  We find it quite
intriguing that the same flow in four dimensions yields two different
solutions to the equations of the 11-dimensional supergravity.

As regards the geometry, we first note that the brane-probe calculation 
for this new solution will be virtually identical to that of the
previous subsection, merely with $\IC\IP^2$ replaced by
$S^2 \times S^2$.   Perhaps more interesting is that 
at the other extreme, {\it i.e.} at $\mu = {\pi \over 2}$, as opposed
to $\mu = 0$, the metric defined using \newframes\ has a conifold 
singularity, and the $S^7$ degenerates to the conifold times $S^1$.  It 
would seem reasonable to conjecture that the solution defined by the metric
based upon $T^{1,1}$ could be related to the T-dual of the compactification
solution of \LJR.  This would imply that it would also be related
to the non-trivial fixed point of the holographic flow of \IKEW.
Adding a flux to this solution, and ellipsoidally squashing would
then represent a flow away from this non-trivial fixed point.  One
might further attempt to generalize this to ADE singularities as
in \SGNNSS, and presumably some of these might be related to using
Einstein-K\"ahler metrics on del Pezzo surfaces.

\subsec{Consistent truncation and simplifying the Ansatz}

In \AKNW\ a number of formulae for the consistent truncation
of the \tobe\  theory were conjectured, and in particular the Ansatz for
$A^{(4)}$ was given in terms of a geometric superpotential.  We find that
we can generalize this directly to M-theory.

We start by converting to the $SL(8,\IR)$ basis, and introducing
rotated vielbeins
$$\eqalign{
U^{ij}{}_{IJ}&\eql u^{ij}{}_{ab}(\Gamma_{IJ})^{ab}\,,
\qquad
V^{ijIJ}\eql v^{ij ab}(\Gamma_{IJ})^{ab}\,,\cr
U_{ij}{}^{IJ}&\eql u_{ij}{}^{ab} (\Gamma_{IJ})^{ab} \,,
\qquad
V_{ijIJ}\eql v_{ijab}(\Gamma_{IJ})^{ab}\,,\cr}
$$
Define:
$$
\eqalign{
A_{ijIJ}& \eql U_{ij}{}^{IJ}+V_{ijIJ}\,,\qquad
B_{ij}{}^{IJ}\eql i(U_{ij}{}^{IJ}-V_{ijIJ})\,,\cr
C^{ij}{}_{IJ}&\eql U^{ij}{}_{IJ}+V^{ijIJ}\,,\qquad
D^{ijIJ}\eql-i( U^{ij}{}_{IJ}-V^{ijIJ})\,, \cr}
$$
and observe that up to a constant, $k$, we have:
$$
T_{l}{}^{kij} \eql k\, C^{ij}{}_{LM}(A_{lmJK}D^{kmKI} \, \delta^L{}_I\,
     \delta^{MJ} -  B_{lm}{}^{JK}C^{km}{}_{KI}\, \delta^{LI}\,
     \delta^M{}_J )\,,
$$
In this expression we have deliberately inserted explicit Kronecker
$\delta$'s as we are going to want to think of $\delta^I_J$ as $SL(8,\IR)$
covariant, but $\delta^{IJ}$ as a metric in a particular $SL(8,\IR)$ frame.
The general idea of \AKNW\ is to introduce geometric analogues of the
$T$-tensor by replacing $\delta^{IJ}$ by $x^Ix^J$, but leaving
$\delta^I_J$ alone.

Now recall that the $A_1$-tensor is defined by
$$
A_1^{ij}\eql T_m{}^{imj}\,.
$$
It is diagonal along the flow considered in section 3, and
indeed the superpotential is read off from  $W \propto A_1^{77}=A_1^{88}$.
Modifying the $A_1$ tensor as outlined above leads to the
geometric $A_1$ tensor, which we will denote by $\widetilde A_1$.
We then find that
$$
\widetilde A_1^{77}+\widetilde A_1^{88}~\propto~ {1 \over 12\, \rho^2 }\,
\big[\, 2  \cosh^2\chi\,\cos^2\mu ~+~ \rho^8\,(3-\cosh(2\chi))\,
\sin^2\mu\, \big] ~\equiv~ \widetilde W,
$$
which is exactly the geometric superpotential introduced in \Athree\ and
\Wtilde.

There are also simplifications that can be made in writing the Ansatz
for $C^{(3)}$.  Define complex coordinates in $\IR^8$ by:
$$
z^1\eql x^1-ix^2\,,\qquad z^2\eql x^3-i x^4\,,\qquad
z^3\eql x^5-i x^6\,,\qquad z^4\eql x^7-ix^8\,.
$$
These coordinates, $z^i$, are non-holomorphic, linear functions of 
the coordinates defined in \spherparam: $z^1 = \cos\mu(u^1-i u^2), 
z^2 = \cos\mu(u^3-i u^4), z^3 = \cos\mu(u^5-i u^6), z^4 =\sin\mu (v^1-i v^2)$. 
Using these new coordinates on $S^7$, we may wrte the internal background
gauge field more simply as:
$$
C_{(3)}\eql { i L^3\over4 \sqrt3} \,{\tanh\chi\over X}\,
 \big(3\,z^{[1}\,dz^2\wedge dz^{3]}\wedge dz^4-\rho^4\,
z^4\,dz^1\wedge dz^2\wedge dz^3\big)\,.
$$

\newsec{Final comments}

We have shown that the solution of \KPNWc\ that represents
the RG flow of $\cN=4$ Yang-Mills theory to the Leigh-Strassler
fixed point can be recast in a more natural form.  In particular,
one sees that this solution represents a significant generalization 
of the class of solutions described in \CNPNPW.  This observation
enables us to generalize easily the supersymmetric flow solution
to $M$-theory.  The Ansatz we obtain is further supported by considering
the T-dual, and lift to $M$-theory, of the \tobe-flow of \KPNWc.
The key ingredients in the Ansatz are to first write the
``round'' compactification in terms of a $U(1)$ fibration
over an Einstein-K\"ahler manifold.  One then ellipsoidally squashes  
the Einstein-K\"ahler base, stretches the $U(1)$ fiber, and 
introduces a tensor gauge potential that proportional to
the holomorphic ``volume form'' on the base.  The latter
tensor is really a non-trivial section of a bundle on the base,
but is globally defined on the total-space of the bundle
provided it is given the proper $U(1)$ charge.
This formulation enables us to immediately generalize our solution
from $S^7$ to manifolds that have del Pezzo spaces as 
sub-manifolds.  In particular we obtain an explicit solution with
a conifold singularity.

The holographic dual of the $\cN=2$ supersymmetric flow in
$M$-theory is an $\cN=2$ supersymmetric flow of the $\cN=8$ 
scalar-fermion theory in  $(2+1)$-dimensions.  This strongly coupled 
theory must therefore have a $\cN=2$ supersymmetric fixed point 
that is completely analogous to the Leigh-Strassler
fixed point of $\cN=4$ Yang-Mills theory.  One might have naively
expected this from field theory in that one can obtain the
$(2+1)$-dimensional theory from trivial dimensional reduction
of $\cN=4$ Yang-Mills theory on a circle.  Alternatively, one
could emulate the arguments of Leigh and Strassler directly 
in $(2+1)$ dimensions.  What makes this less obvious is that 
the $(2+1)$-dimensional theory is strongly coupled even at the 
UV fixed point.  Moreover, trivial dimensional reduction of
the Yang-Mills theory will yield a theory at a different point in
the Coulomb branch moduli space compared to the holographic 
UV fixed point in $M$-theory:  As we saw, reduction and T-dualization
leads to a uniform distribution of branes on a torus, and not to
a set of localized branes.   It was not {\it a priori} clear
that this difference in moduli would be consistent with the
flows to non-trivial fixed points.  In terms of the gauged 
supergravity,  the potentials of gauged four- and five-dimensional
supergravity are rather different, and it was far from clear
that a flow in one theory would be directly convertible 
into a flow in the other.  Our results show that, at least for
these supersymmetric flows, the naive expectations are in fact correct.
It would be very interesting to probe the extent of this 
correspondence:  The indications from field theory \refs{\NDorey,
\OANDSPK} are very favorable, at least for the ground state
structure and domain walls.  The supergravity version of
this story is somewhat murkier, but should be clarified in the 
near future \dWNWnew.

There are now many solutions of \tobe\ supergravity that
correspond to RG flows in four-dimensional field theories.
For those solutions constructed directly in ten dimensions,
the initial Ansatz and the final solution usually exploit
details of some non-trivial topology, such as a non-trivial
cycle and a non-trivial flux upon it.  On the other hand,
flows generated by masses or vevs from a maximally
supersymmetric field theory, such as the one considered
here or those of \refs{\FGPWb,\KPNWb,\KPNWc}, generally
have no non-trivial topology to exploit.  As a result, the
ten- or eleven-dimensional geometry is rather hard to
characterize. The results here at least shed a little more 
light on the geometric structure  of these ``non-topological'' flows.  
One feels that the half-maximal supersymmetric flows (such as
that of \KPNWb) should have a particularly simple geometric
characterization.  So far this has eluded us, but such a 
characterization could be very useful: As was shown in 
\refs{\BPP,\EJP} the $\cN=2$ flow of \KPNWb\ represents
only one point on the continuum moduli space of the large $N$,
Seiberg-Witten effective action.  It would be very interesting
to find the general solution with all the moduli, and understanding
the supergravity geometry is probably crucial to doing this.

\bigskip
\leftline{\bf Acknowledgements}
We would like to thank Jaume Gomis for many discussions, and 
Clifford Johnson for his comments and for giving us a preliminary
copy of \CVJnew. This work was supported in part by funds provided
by the DOE under grant number DE-FG03-84ER-40168.


\listrefs
\vfill
\eject
\end